\documentclass[acmsmall]{acmart}

\usepackage{pifont}
\usepackage[most]{tcolorbox}
\usepackage[normalem]{ulem}
\usepackage{threeparttable}
\usepackage{enumitem}
\usepackage{csquotes}
\usepackage{multirow}
\usepackage{url}
\usepackage{hyperref}
\usepackage{fontawesome5} % 提供 \faLink 图标
\usepackage{algorithm}
\usepackage{algpseudocode}
\usepackage{subcaption}
\usepackage{makecell}
\usepackage{placeins}

\def\eg{\textit{e.g.,} }
\def\ie{\textit{i.e.,} }

\newcommand{\ours}[1]{\textsc{ReqElicitGym}}

\tcbset{
  colframe=black!10,
  colback=white,
  coltitle=black,
  fonttitle=\bfseries,
  boxrule=0.4pt,
  left=6pt,right=6pt,top=6pt,bottom=6pt,
  title style={left color=black!3,right color=black!1}
}

\AtBeginDocument{%
  }
\newtcolorbox{boxK}{
    top=2pt,
    bottom=2pt,
    left=2pt,
    right=2pt,
    boxrule = 0pt,
    toprule = 0pt, % top rule weight
    colback=gray!15,   % 浅灰背景色（数字越小越浅，可调 5~20）
    colframe=white     % 去掉边框颜色，和 boxrule=0pt 配合
}

\AtBeginDocument{%
  }

\setcopyright{acmlicensed}
\copyrightyear{2025}
\acmYear{2025}
\acmDOI{}
\acmJournal{TOSEM}
\acmISBN{}

\begin{document}


\title{\ours{}: An Evaluation Environment for Interview Competence in Conversational Requirements Elicitation}

% \title{\ours{}: An Automatic Evaluation Environment for Requirements Elicitation Interview}

% Automatically Benchmarking the Interview Competence of LLMs for Conversational Requirements Elicitation}


\author{Dongming Jin}
\email{dmjin@stu.pku.edu.cn}
\affiliation{%
  \institution{Peking University}
  \city{Beijing}
  \country{China}
}

\author{Zhi Jin}
\email{zhijin@pku.edu.cn}
\authornote{Zhi Jin is the Corresponding author}
\affiliation{
  \institution{Peking University and Wuhan University}
  \city{Beijing}
  \country{China}
}

\author{Zheng Fang}
\affiliation{%
  \institution{Peking University}
  \city{Beijing}
  \country{China}
}

\author{Linyu Li}
\affiliation{%
  \institution{Peking University}
  \city{Beijing}
  \country{China}
}

\author{XiaoTian Yang}
\affiliation{%
  \institution{Beijing Forest University}
  \city{Beijing}
  \country{China}
}

\author{Yuanpeng He}
\affiliation{%
  \institution{Peking University}
  \city{Beijing}
  \country{China}
}

\author{Xiaohong Chen}
\affiliation{
  \institution{East China Normal University}
  \city{Shanghai}
  \country{China}
}

\renewcommand{\shortauthors}{Jin et al.}

%%
%% The abstract is a short summary of the work to be presented in the article.
\begin{abstract}
With the rapid improvement of LLMs' coding capabilities, the bottleneck of LLM-based automated software development is shifting from generating correct code to eliciting users' requirements. Despite growing interest, the interview competence of LLMs in conversational requirements elicitation remains fully underexplored. Existing evaluations often depend on a few scenarios, real user interaction, and subjective human scoring, which hinders systematic and quantitative comparison. To address these challenges, we propose \ours{}, an interactive and automatic evaluation environment for assessing interview competence in conversational requirements elicitation. Specifically, \ours{} introduces a new evaluation dataset and designs both an interactive oracle user and a task evaluator. The dataset contains 101
website requirements elicitation scenarios spanning 10 application types. Both the oracle user and the task evaluator achieve high agreement with real users and expert judgment. Using our \ours{}, any automated conversational requirements elicitation approach (\eg LLM-based agents) can be evaluated in a reproducible and quantitative manner through interaction with the environment. 
% We further conduct a systematic empirical study on seven representative LLMs
Based on our \ours{}, we conduct a systematic empirical study on seven representative LLMs, and the results show that current LLMs still exhibit limited interview competence in uncovering implicit requirements. Particularly, they elicit less than half of the users' implicit requirements, and their effective elicitation questions often emerge in later turns of the dialogue. Besides, we found LLMs can elicit  \textit{interaction} and \textit{content} implicit requirements, but consistently struggle with \textit{style}-related requirements. We believe \ours{} will facilitate the evaluation and development of automated conversational requirements elicitation, and we release our dataset and code at https://github.com/jdm4pku/ReqElicitBench.
\end{abstract}





\received{20 February 2007}
\received[revised]{12 March 2009}
\received[accepted]{5 June 2009}

\maketitle

\section{Introduction}

\begin{displayquote}
\textsf{A problem well stated is a problem half solved.}  
\begin{flushright}
\footnotesize ------ Charles Kettering
\end{flushright}
\end{displayquote}

Large Language Models (LLMs), such as OpenAI's GPT~\cite{achiam2023gpt} and Google's Gemini~\cite{comanici2025gemini}, have demonstrated significant capabilities in code generation, enabling an emerging wave of LLM-based automated software development. As the coding capability improves, the primary bottleneck of LLM-based automated software development is gradually shifting from generating syntactically and semantically correct code to eliciting accurate and complete user requirements~\cite{jin2025iredev}. In practice, many software failures stem not from incorrect implementations but from misunderstandings and omissions during requirements elicitation~\cite{jones2004software}~\cite{christel1992issues}. \textbf{Thus, conducting effective requirements elicitation has become increasingly important in the era of AI-assisted software development.}

\begin{figure}[htbp]
    \centering
 \includegraphics[width=0.99\linewidth]{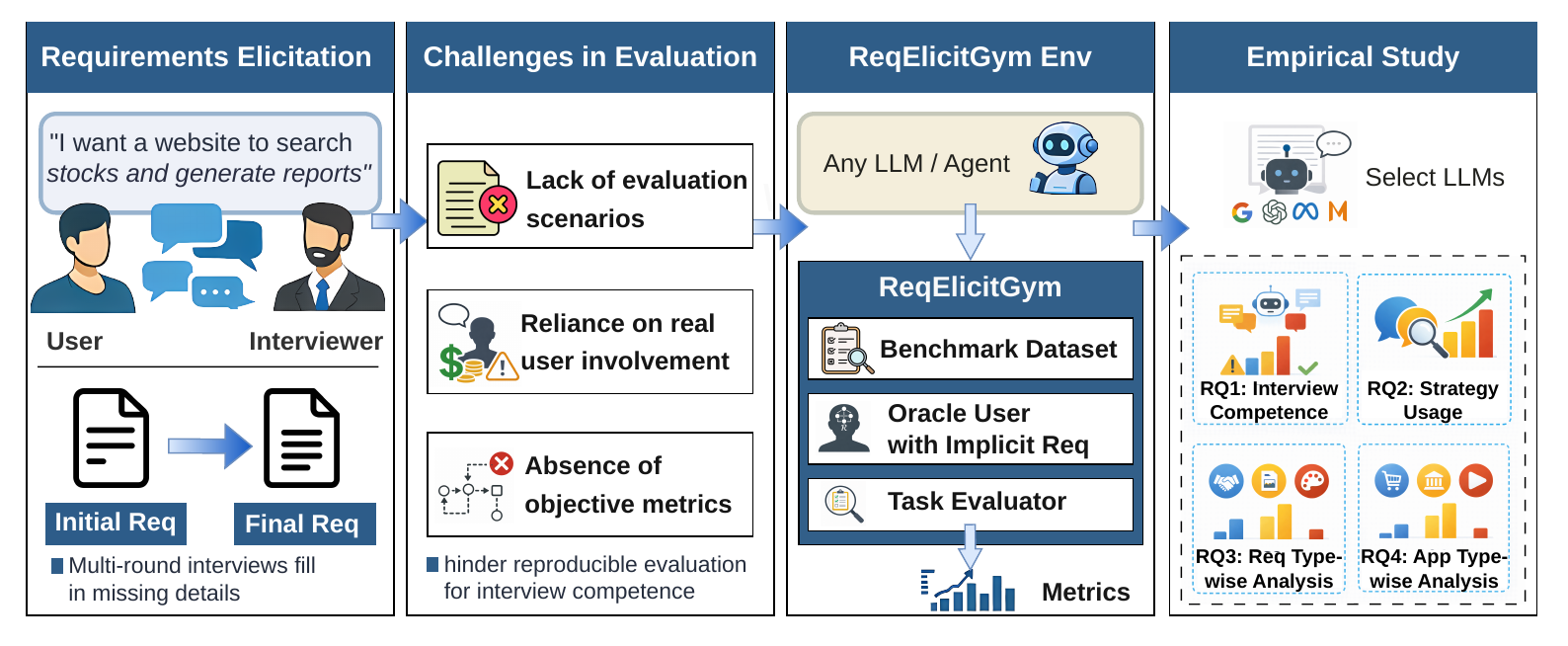}
    \caption{End-to-end illustration of the problem, challenges, evaluation environment, and empirical study for conversational requirements elicitation.}
    \label{fig:teaser}
\end{figure}

% 对话式需求获取是一种常见的方式。
Conversational requirements elicitation~\cite{rietz2019designing} is a common way to refine an initial and underspecified requirements description into a complete requirements description through multi-turn interviews. As shown in Figure~\ref{fig:teaser}, an interviewer (\eg a product manager) conducts a multi-turn conversation with stakeholders to clarify goals, uncover implicit requirements, and resolve ambiguities. For example, an initial stakeholder requirements description might be ``\textit{I want a website that allows me to search for stocks and generate reports}''. This description leaves many details implicit (\eg search mechanisms, report contents, and output format). The interviewer must ask targeted clarification questions to elicit the missing information.

Recently, researchers have begun exploring automated support for conversational requirements elicitation. However, most of the studies~\cite{almeida2025elicitation}~\cite{ataei2025elicitron}~\cite{jin2025iredev} focus on evaluating the quality of requirements generated in a single step, while lacking the evaluation of interview competence during multi-turn conversational requirements elicitation. To the best of our knowledge, only one closely related work~\cite{KornGV25} has explored the feasibility of conducting multi-turn conversational requirements elicitation using LLMs. Specifically, they proposed an LLM-based requirements elicitation chatbot and conducted human-in-the-loop interview experiments with students across two development scenarios.

% The limitation is that their evaluation was limited to only two scenarios and relied heavily on manual participation and subjective assessment. 

Thus, LLMs' interview competence still remains fully underexplored. This is due to the following challenges in current evaluation practice. \textbf{(1) Lack of evaluation scenarios.} Existing evaluations~\cite{KornGV25} typically cover only a small number of scenarios, which makes it difficult to support systematic comparison. \textbf{(2) Reliance on real user involvement.} Current evaluations~\cite{KornGV25}~\cite{ferrari2019learning}~\cite{ferrari2020sapeer} largely depend on real user involvement, \ie human role-played stakeholders or student participants. This requires substantial human cost and introduces variability due to differences in user behavior and engagement. \textbf{(3) Absence of objective and process-aware metrics.} Prior work~\cite{KornGV25}~\cite{ferrari2020sapeer} often relies on post-hoc subjective ratings and lacks metrics that track what requirement information is elicited across turns, hindering quantitative and reproducible evaluation.

To address these challenges, we propose \ours{}, an interactive and automatic evaluation environment for assessing interview competence in conversational requirements elicitation. Using \ours{}, any automated conversational requirements elicitation approach (\eg, LLM-based agents) can be evaluated in a reproducible and quantitative manner through direct interaction with the environment. Specifically, \ours{} introduces a new evaluation dataset and designs both an interactive oracle user and a task evaluator. The dataset contains 101 requirements elicitation scenarios spanning 10 application domains. As shown in Figure~\ref{fig:dataset_overview}, the core design philosophy of this dataset is to explicitly capture the requirements gap (\ie Implicit Req) between an underspecified initial requirements description (\ie Initial Req) and a complete final requirements description (\ie Final Req). The oracle user is played by GPT-5.1 under controlled prompts to simulate a realistic stakeholder. Its responses to the interviewer’s questions are grounded in predefined implicit requirements. This enables large-scale and reproducible evaluation without the involvement of real users. The task evaluator continuously tracks the progress of requirement elicitation. Based on an LLM-as-a-Judge mechanism, it can determine whether and which implicit requirements are successfully elicited at each conversation turn. At the end of the interview, it computes a series of metrics (\eg Implicit Requirement Elicitation Rate) to evaluate elicitation effectiveness. In addition, it also supports process-aware analysis across dialogue turns. We validate both the oracle user and task evaluator using 33 real requirements interview dialogues, comprising a total of 555 turns. The oracle user obtained a Cohen’s $\kappa$ score\footnote{Following the widely adopted interpretation proposed~\cite{landis1977measurement}, Cohen’s $\kappa$ values above 0.60 are generally regarded as indicating substantial agreement.} of 0.73 against real users in terms of implicit requirement disclosure behavior. The task evaluator obtained a Cohen’s $\kappa$ score of 0.72 against expert judgments, indicating that it can reliably determine whether implicit requirements are elicited at each conversation turn. 

Finally, we conduct a systematic empirical study based on \ours{} to investigate the interview competence of representative mainstream LLMs. We evaluate seven recent LLMs under two inference settings (\ie Non-CoT and CoT) using a unified experimental protocol. Our study examines not only overall elicitation effectiveness (IRE), but also questioning efficiency (TKQR), strategy usage (ESR), and requirement-type coverage. Our results reveal several interesting findings. First, current mainstream LLMs still exhibit limited interview competence. Even the best-performing model achieves an IRE of only 0.32, indicating that a substantial portion of implicit requirements cannot be systematically uncovered. Second, while CoT prompting consistently improves questioning efficiency (TKQR) and reduces dialogue turns, it does not reliably increase elicitation coverage (IRE). Third, LLMs overwhelmingly favor probing over clarification, with probing turns far exceeding clarification turns, while clarification questions demonstrate consistently lower effectiveness (ESR). Besides, clear disparities exist across requirement types: Interaction and Content requirements are moderately elicitable, whereas Style-related requirements remain near zero, revealing a structural weakness in uncovering subjective user preferences. These findings highlight a significant gap between the current conversational capabilities of LLMs and the systematic reasoning required for effective real-world requirements elicitation.

In summary, this paper makes the following contributions:
\begin{itemize}
    \item We propose \ours{}, the first automated and reproducible evaluation environment for assessing interview competence in conversational requirements elicitation. 
    \item \ours{} introduces an evaluation dataset with 101 requirements elicitation scenarios spanning 10 application types, explicitly modeling \textit{initial req}, \textit{implicit req}, and \textit{final req}.
    \item \ours{} designs an oracle user and a task evaluator to support automatic and consistent evaluation without real user participation.
    \item We conduct a systematic empirical study on representative LLMs using \ours{}, providing quantitative analysis of their interview competence.
\end{itemize} 

\section{Background and Related Works}

\subsection{Requirements Elicitation Interviews}

Requirements elicitation has long been recognized as a critical and challenging phase in software engineering, with many project failures attributed not to technical implementation issues but to deficiencies in understanding and capturing stakeholders’ needs~\cite{jones2004software}. Among various elicitation techniques, interviews are one of the most widely used and effective approaches~\cite{ferrari2016ambiguity}, as they enable analysts or product managers to directly interact with stakeholders and explore their expectations, preferences, and constraints. Christel and Kang et al~\cite{christel1992issues} described a typical interview process consisting of four main stages: \emph{(1) preparation}: identify stakeholders and defining interview goals; \emph{(2) interviewing}: elicit information from stakeholders; \emph{(3) recording}: document elicited information; and \emph{(4) integration}: synthesize requirements from collected information.

Prior work has proposed various methods and tools to support effective human-conducted requirements elicitation interviews across these stages. For the preparation stage, Zaremba et al. ~\cite{zaremba2021towards} proposed a systematic typology of questions used in the interviewing stage. For the interviewing stage, several studies focus on improving human interview skills through training and education, including learning from common interviewer mistakes~\cite{bano2019teaching} and role-based pedagogical approaches such as role-playing, self-assessment, peer review, and role reversal~\cite{ferrari2019learning,ferrari2020sapeer}. To improve scalability in interview training, G\"orer and Aydemir~\cite{gorer2024exploring} proposed the REIT architecture based on robotic and virtual tutors. For the recording stage, Debnath et al.~\cite{debnath2022annoterei} introduced AnnoteREI to facilitate the transcription and annotation of interview data. For requirement integration, Voria et al.~\cite{voria2025recover} proposed an automated framework to generate system requirements from stakeholders’ conversations.  In addition, other studies further investigated intrinsic characteristics of requirements interviews, such as the role of ambiguity in eliciting tacit knowledge~\cite{ferrari2016ambiguity} and the impact of analysts’ domain knowledge on interview effectiveness~\cite{hadar2014role}.

Recently, with the rapid advancement of large language models, researchers have begun exploring intelligent and automated support for requirements elicitation. One category of studies investigate the use of LLMs to generate interview scripts~\cite{gorer2023generating}.  Another category explore using LLMs to extract or generate system requirements directly from interview dialogues, aiming to reduce the cost of manual requirement consolidation~\cite{almeida2025elicitation}. These studies demonstrate that LLMs exhibit promising capabilities in requirement preparation and requirements integration. Despite these advances, the interview competence itself (\eg ask targeted questions) remains insufficiently and inconsistently evaluated. Existing studies focus on evaluating the quality of requirements generated in a single step, while lacking evaluation of interview competence during multi-turn conversational requirements elicitation. To the best of our knowledge, only one closely related work~\cite{KornGV25} has explored the feasibility of conducting conversational requirements elicitation using LLMs. However, their evaluation was limited to only two scenarios and relied heavily on manual participation and subjective assessment. Thus, this paper aims to construct an automatic and reproducible environment for assessing interview competence in conversational requirements elicitation, facilitating systematic comparison and further development of automated methods. 

\subsection{Proactive Task Benchmark}
Traditional benchmarks~\cite{chen2021evaluating}~\cite{jain2024livecodebench} primarily evaluate performance under fully specified prompts, assuming that user intentions are clearly and completely articulated. Such benchmarks mainly assess whether the model can generate correct outputs when sufficient information is provided. However, in real-world human–AI interaction scenarios, this assumption often does not hold. Users frequently express their needs in incomplete, incremental, or ambiguous ways. Motivated by this gap, recent research has begun to shift toward proactive and interactive task benchmarks.

Recent work increasingly emphasizes user-centered evaluation settings. Wang et al.~\cite{wang2024user} constructed a benchmark highlighting intention-aware evaluation and ecological validity. Similarly, WildBench~\cite{lin2024wildbench} selects challenging tasks from authentic user interaction logs, observing that model performance in real-world environments is significantly lower than that reported on academic benchmarks. Beyond static evaluation, several benchmarks focus on multi-turn interaction and tool collaboration. MINT~\cite{wang2023mint} evaluates a model’s ability to iteratively revise task solutions through tool invocation and language feedback in multi-turn settings. $\tau$-bench~\cite{yao2024tau} and its extension $\tau^2$-Bench~\cite{barres2025tau} construct structured environments to assess long-horizon reasoning, state tracking, and decision robustness in tool–agent–user interactions. In addition, some research examines user-centered and personalized interaction capabilities. UserBench~\cite{qian2025userbench} provides an interactive environment simulating evolving user preferences and satisfaction to evaluate long-term adaptation ability. Overall, current evaluation paradigms are evolving from isolated task completion toward comprehensive assessment of interaction quality and user adaptation.

Despite these advances, existing proactive and interactive benchmarks primarily focus on task completion, tool collaboration, or preference alignment. They rarely address a crucial prerequisite stage of real-world task execution: requirements elicitation.  In many practical domains (\eg software development), the user’s initial request is inherently incomplete. Systems must proactively uncover implicit requirements before proceeding to implementation. Unlike tool-driven benchmarks that assume clearly defined task objectives, conversational requirements elicitation emphasizes identifying missing constraints, clarifying ambiguous intentions, and progressively transforming an underspecified problem statement into a complete requirements specification through multi-turn interaction. 

To the best of our knowledge, there is currently no structured and reproducible benchmark specifically designed to evaluate the interview competence of conversational agents in requirements elicitation. Our proposed \ours{} builds upon the research trajectory of proactive task benchmarking but focuses explicitly on the pre-execution stage. It evaluates whether an agent can effectively identify requirement gaps, pose targeted clarification and probing questions. In this sense, \ours{} extends the evaluation boundary from task-solving ability to problem-formulation ability, which is a foundational capability for reliable AI-assisted software development.

\section{ReqElicitGym Environment}
In this section, we present \ours{}, an automated and reproducible evaluation environment for assessing interview competence in conversational requirements elicitation. We first define the overview of our \ours{} and describe the details in the following sections, including three key modules and the implementation details.

\subsection{Overview}
The goal of \ours{} is to provide a controlled, scalable, and reproducible evaluation environment for measuring an interviewer’s ability to elicit missing and implicit requirements through multi-turn dialogue. To achieve this goal, \ours{} decomposes the evaluation task into three modules: an evaluation dataset, an oracle user, and a task evaluator. These three modules work in a pipeline as shown in Figure~\ref{fig:env}:

\begin{figure}[htbp]
    \centering
\includegraphics[width=0.99\linewidth]{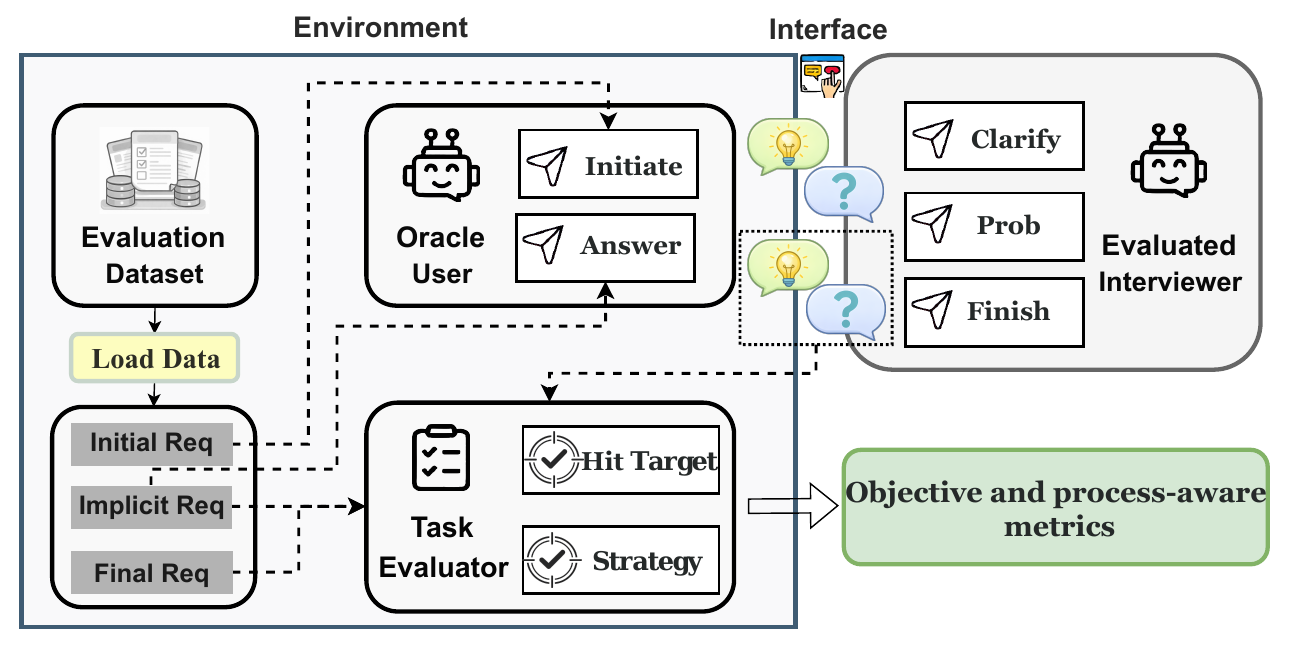}
    \caption{Overview of the \ours{} evaluation environment.}
    \label{fig:env}
\end{figure}

\begin{itemize}
    \item \textbf{Evaluation Dataset:} provides a series of requirements elicitation scenarios. Each scenario acts as an elicitation task and consists of three core components, \ie an \textit{initial req}, a set of ground-truth \textit{implicit req}, and a \textit{final req}.
    \item \textbf{Oracle User:} simulates a realistic stakeholder by generating responses to the interviewer’s questions and revealing implicit requirements in a controlled and principled manner.
    \item \textbf{Task Evaluator:} analyzes the multi-turn dialogue between the interviewer and the oracle user, and quantitatively evaluates the effectiveness of the elicitation process.
\end{itemize}

\textbf{Environment Execution Process.}
For each elicitation scenario in the evaluation dataset, the oracle user initiates the interaction by presenting the \textit{initial req}. The interviewer then conducts a multi-turn dialogue by iteratively selecting a strategy to pose questions until it decides to terminate the interview. At each turn, the oracle user generates responses grounded in the scenario’s implicit requirements. Meanwhile, the task evaluator examines the dialogue turn by turn to determine whether the interviewer has successfully elicited implicit requirements and judges the strategy the interviewer adopts. Based on these turn-level judgments, \ours{} computes a set of quantitative evaluation metrics that characterize the interviewer’s performance. The detailed definitions of these metrics are presented in Section~\ref{subsec:metrics}.

\subsection{Evaluation Dataset}

\subsubsection{Dataset Overview.} 

\begin{figure}[htbp]
    \centering
    \includegraphics[width=0.99\linewidth]{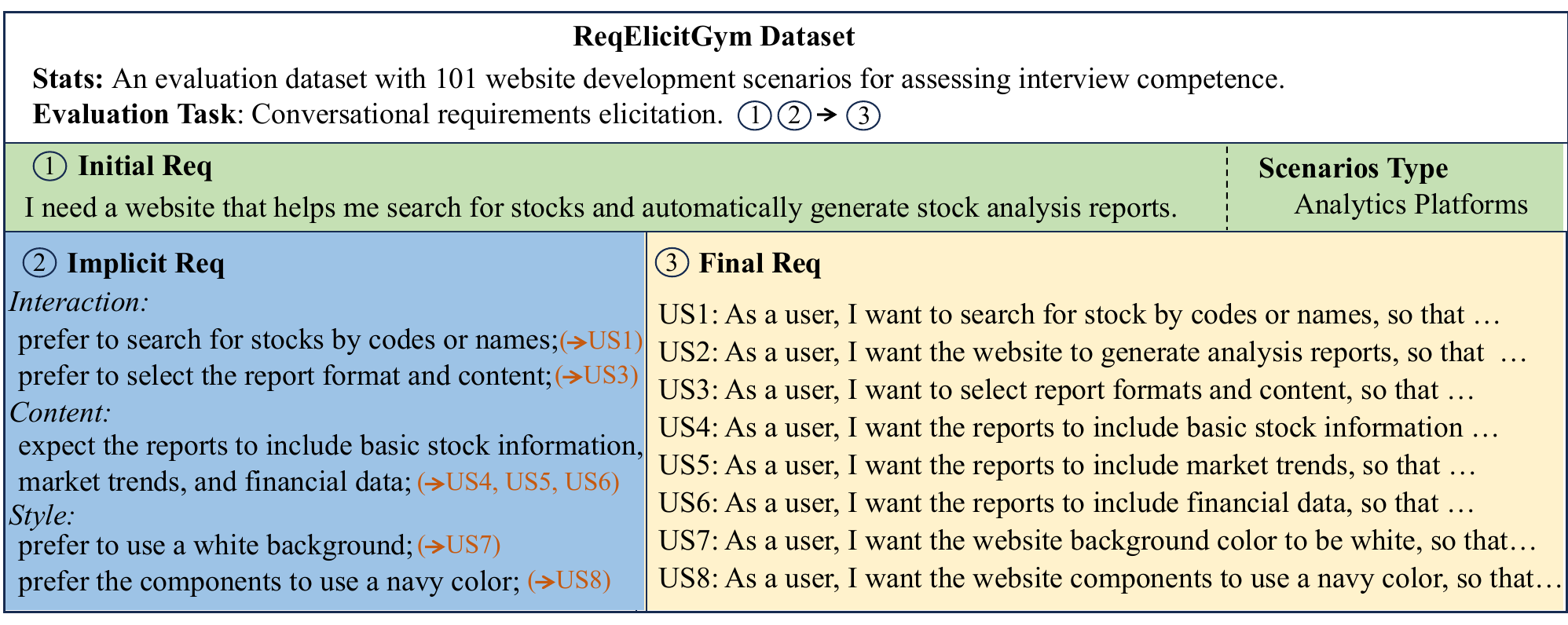}
    \caption{Structure of a requirements elicitation scenario in the evaluation dataset.}
    \label{fig:dataset_overview}
\end{figure}

The dataset focuses on website development scenarios and contains 101 website requirements elicitation scenarios spanning 10 application types. As shown in the Figure~\ref{fig:dataset_overview}, each scenario consists of three core components. \ding{182} \textbf{Initial Req}: represents a realistic user requirement description provided at the early stage of a software project. It typically captures the core functionalities that users care about, but omits many critical details necessary for implementation. \ding{183} \textbf{Implicit Req}: consists of a set of critical but missing details that are not explicitly stated in the initial spec but must be elicited through conversation. Each implicit requirement is categorized into one of three dimensions, \ie user interaction, content representation, or visual style. \ding{184} \textbf{Final Req}: provides a complete and well-defined requirements description obtained after all implicit requirements have been successfully elicited and confirmed through interviews. 

\subsubsection{Dataset Construction Pipeline.}
% Figure~\ref{} illustrates the procedure of constructing our dataset. 
We carefully follow five steps to construct it. (i) software project collection; (ii) annotation preparation; (iii) final spec annotation; (iv) initial spec annotation; (v) implicit req annotation; (vi) quality control and interactive refinement.

\textbf{(i) Software Project Collection.}  The software projects used in this dataset are derived from an existing website development benchmark, \ie WebGen-Bench~\cite{lu2025webgen}. Specifically, we select its evaluation dataset, which contains 101 realistic website system development scenarios covering a diverse range of typical web application types. For each development scenario, WebGen-Bench provides an English instruction that describes a relatively complete website development requirement, including both functional requirements and non-functional constraints such as visual style. These instructions serve as a high-quality and realistic source of requirements for constructing our dataset. In addition, we chose website development as the target scenario for requirements elicitation for three main reasons. First, website development is highly representative of real-world software engineering practice, as web systems are among the most common and widely deployed types of software applications. Second, web projects at their early stages are often characterized by highly under-specified requirements. Users typically provide only high-level intent descriptions, while many critical decisions (\ie information content, interaction behaviors, and visual preferences) must be gradually clarified. Third, web applications inherently involve both functional and non-functional requirements, making them particularly suitable for studying the elicitation of diverse types of implicit requirements.

\textbf{(ii) Annotation preparation.} 
Before starting the annotation process, we carried out three preparatory efforts to ensure annotation quality and consistency. \textit{(1) Developing a annotation platform.} we developed a dedicated annotation platform to support the efficient annotation and revision of core components in our dataset. The platform presents the original instruction description for each scenario and allows annotators to annotate and edit core components, including Initial Req, Implicit Req, and Final Req. \textit{(2) Recruiting qualified annotators.} We recruited three Ph.D. candidates from a prestigious university. Each annotator has over four years of SE experience, fundamental knowledge of requirements elicitation, and an participation of multiple website development projects. This background enables them to annotate Initial Spec and Implicit Req and Final Spec accurately. \textit{(3) Conducting unified annotator training.} We conduct a unified training process to ensure a consistent understanding of the annotation task and criteria. The training included detailed annotation guidelines that clearly defined the scope, objectives, and annotation principles of the three core components along with representative examples for each type. During this phase, annotators conducted pilot annotations and discussed the results collaboratively to resolve ambiguities, align standards, and refine annotation practices. 

\textbf{(iii) Final Spec Annotation.}
For each development scenario, annotators first annotated the Final Spec, as it is the most well-defined and controllable component in our dataset. 
% The Final Spec serves as the reference complete specification against which the requirements elicitation process is evaluated.
Specifically, annotators rewrote the original website development instruction into a complete and structured requirements specification expressed as a set of user stories. Each user story describes a concrete requirement in a clear and unambiguous manner, covering aspects such as system functional behavior, content presentation requirements, or visual and stylistic constraints. Together, these user stories form a comprehensive specification that fully captures the intended system behavior and constraints implied by the original instruction.

\textbf{(iv) Initial Spec Annotation.}
Based on the annotated Final Spec, annotators further constructed the corresponding Initial Spec for each development scenario. The key challenge in this step lies in carefully controlling the scope and degree of requirement under-specification, balancing realism and elicitation rooms. To alleviate this challenge, the annotators followed this explicit guideline: the Initial Spec should preserve all core functional intents of the software project while deliberately omitting details that would typically require further clarification during requirements interviews. For example, an Initial Spec may state that the system supports search functionality without specifying search criteria, or mention information presentation features without detailing which fields should be displayed. We acknowledge that this construction may reduce the overall difficulty of the requirements elicitation task to some extent. The associated design trade-offs and their potential impact are discussed in detail in Section~\ref{subsec:threats}.

% application type组合。

\textbf{(v) Implicit Req Annotation.}
After constructing the Initial Spec and Final Spec for each development scenario, annotators further constructed the corresponding Implicit Requirements. The goal of this step is to explicitly capture the missing information that must be elicited through conversation in order to transform an initial Spec into a complete final Spec. Specifically, annotators performed a systematic comparison between the user stories in the Final Spec and their corresponding descriptions in the Initial Spec. Any information that is present in the Final Spec but is absent or underspecified in the Initial Spec was annotated as an Implicit Requirement. To enable fine-grained analysis, each implicit requirement was further categorized into one of three dimensions according to its nature, \ie user interaction (\eg interaction flows and user operations), content representation (\eg information fields and content organization), or visual style (\eg layout and aesthetic preferences). This categorization reflects common characteristics of website requirements and facilitates a structured evaluation of eliciting different types of implicit information.

\textbf{(vi) Quality Control and Iterative Refinement.}
To ensure the overall quality and internal consistency of the dataset, we conducted multiple rounds of quality control and iterative refinement. The primary criterion for quality validation is that the set of Implicit Requirements must fully bridge the information gap between the Initial Spec and the Final Spec, without introducing redundant, irrelevant, or speculative details. After each annotation round, the first author performed a comprehensive review of all scenarios, focusing on identifying three types of issues: missing implicit requirements necessary to derive the Final Spec from the Initial Spec, overly specific Initial Specs that prematurely disclose details intended to be elicited through conversation, and inconsistencies among the Initial Spec, Implicit Requirements, and Final Spec. Scenarios containing such issues were flagged and revised in subsequent iterations. This process was repeated for three rounds, during which the number of identified issues consistently decreased (from 26 to 9 to 0), indicating convergence of annotation quality. After the final iteration, all scenarios passed the quality checks, resulting in a dataset with high reliability and strong internal consistency.

\subsubsection{Dataset Statistics}

\begin{figure}[t]
    \centering
    \begin{subfigure}[b]{0.4\textwidth}
        \centering
        \includegraphics[width=\textwidth]{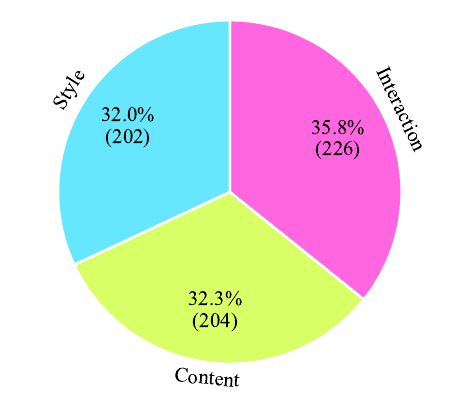}
        \caption{Distribution of implicit requirement types across all scenarios in the evaluation dataset.}
        \label{fig:implicit_type_distribution}
    \end{subfigure}
    \hfill
    \begin{subfigure}[b]{0.4\textwidth}
        \centering
        \includegraphics[width=\textwidth]{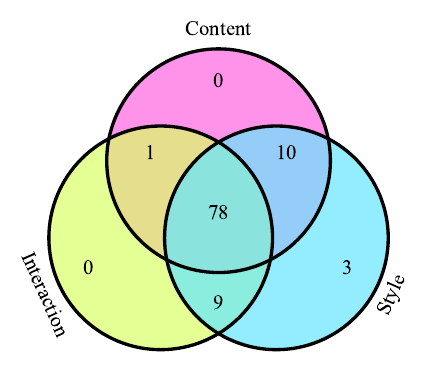}
        \caption{Distribution of scenarios with respect to different combinations of implicit requirement categories.}
        \label{fig:implicit_combination_distribution}
    \end{subfigure}
    \caption{Statistics of implicit requirements in the evaluation dataset.}
    \label{fig:two_images}
\end{figure}

Table~\ref{tab:dataset_statistics} summarizes the overall statistics of our evaluation dataset. 
The dataset consists of 101 requirements elicitation scenarios spanning 10 application types, covering a wide range of realistic web-based systems.
Across all scenarios, we annotate a total of 101 \textit{Initial Req}, 632 \textit{Implicit Req} and 1,000 \textit{Final Req}. Figure~\ref{fig:implicit_type_distribution} illustrates the distribution of implicit requirement types in the dataset. The three implicit requirement categories exhibit a relatively balanced distribution, with Interaction being slightly more prevalent. Figure~\ref{fig:implicit_combination_distribution} further presents the distribution of scenarios with respect to different combinations of implicit requirement categories.
The results show that most scenarios contain implicit requirements across all three aspects, while scenarios involving only a subset of aspects account for a much smaller proportion of the dataset. Overall, the dataset contains a sufficient number of evaluation scenarios, covers a diverse range of application types, and reflects the multifaceted nature of implicit requirements in real-world scenarios.

\begin{table}[]
    \centering
    \setlength{\tabcolsep}{3pt}
    \caption{Statistics of the evaluation dataset across different application types. The table reports the number of scenarios (\#Sce) per application type, along with the minimum, average, and maximum lengths of initial requirements, the number of annotated implicit requirements, and the number of final requirements.}
    \begin{tabular}{lcccccccccc}
\toprule
\multirow{2}{*}{\textbf{Application Type}} & \multirow{2}{*}{\textbf{\#Sce}} & \multicolumn{3}{c}{\textbf{Length of Initial Req}} & \multicolumn{3}{c}{\textbf{\#Implicit Req}} & \multicolumn{3}{c}{\textbf{\#Final Req}}   \\
                                           &  & \textbf{Min} & \textbf{Avg} & \textbf{Max} & \textbf{Min} & \textbf{Avg} & \textbf{Max} & \textbf{Min} & \textbf{Avg} & \textbf{Max} \\ \midrule
Showcase Websites                          & 22 & 9  & 29.88 & 39 & 2 & 4.25 & 7  & 5 & 8.64 & 12 \\
Community Platforms                        & 12 & 7  & 14.75 & 22 & 2 & 4.75 & 6  & 4 & 7.67 & 11 \\
E-commerce Web                             & 16 & 13 & 20.75 & 33 & 3 & 5.25 & 8  & 7 & 10.44 & 17 \\
Learning Platforms                         & 8  & 19 & 21.38 & 30 & 5 & 8.38 & 12 & 7 & 10.25 & 18 \\
Entertainment App                          & 7  & 11 & 27.43 & 38 & 2 & 4.43 & 6  & 6 & 12.43 & 18 \\
Dashboards                                 & 6  & 13 & 19.33 & 26 & 4 & 5.00 & 6  & 8 & 8.67 & 10 \\
Enterprise Management                      & 8  & 16 & 21.33 & 27 & 4 & 7.17 & 9  & 6 & 14.00 & 24 \\
Publishing Platforms                       & 4  & 17 & 19.00 & 22 & 3 & 5.50 & 8  & 5 & 9.75 & 15 \\
Job Search Platforms                       & 4  & 16 & 18.00 & 21 & 7 & 9.25 & 11 & 9 & 11.25 & 14 \\
Productivity Tool                          & 14 & 14 & 17.40 & 21 & 3 & 7.60 & 10 & 6 & 9.57 & 14 \\ \midrule
\textbf{Total}                             & \textbf{101} & \textbf{7} & \textbf{21.86} & \textbf{39} & \textbf{2} & \textbf{6.26} & \textbf{12} & \textbf{4} & \textbf{9.90} & \textbf{24} \\
\bottomrule
\end{tabular}
    \label{tab:dataset_statistics}
\end{table}

\FloatBarrier

\subsection{Oracle User}

\subsubsection{Construction.}

To avoid heavy reliance on real user participation and reduce variability introduced by different users across experiments, \ours{} introduces an LLM-based oracle user to simulate realistic interviewees in conversational requirements elicitation. The oracle user serves as a controlled and reproducible proxy for human stakeholders, responding to interview questions in a manner that is consistent with predefined implicit requirements.

The construction of the oracle user follows three key design principles. \textbf{(1) Groundedness in Implicit Requirements.} The oracle user must only reveal information that is explicitly annotated as implicit requirements in the evaluation dataset. Its responses are grounded in the predefined \textit{Implicit Req} set of each scenario, ensuring that no information beyond the ground truth is introduced during the interaction. \textbf{(2) Passive Response.} The oracle user adopts a controlled and non-proactive interaction style to enable fair and discriminative evaluation. Specifically, it does not proactively volunteer additional requirement information unless explicitly asked by the interviewer. This design isolates the interviewer’s contribution and more accurately reflects its ability to identify missing requirement dimensions and formulate effective questions. \textbf{(3) Context Awareness.} The oracle user conditions its responses on the full dialogue history, allowing it to maintain consistency across turns and avoid repeating previously disclosed information. Figure~\ref{fig:oracle_prompt} presents the detailed prompt design of the oracle user, which operationalizes the above principles through explicit role definition, response constraints, and structured conversational context.

\begin{figure}[t]
\centering
\begin{tcolorbox}[
    sharp corners,
    width=0.95\textwidth,
    colback=gray!5,
    colframe=black,
    boxrule=0.6pt,
    fontupper=\footnotesize,
    title=\textbf{Oracle User Prompt Design},
    before skip=4pt,
    after skip=0pt
]

\textbf{System Role.}
You are a user being interviewed about your software requirements.
Respond naturally to the interviewer's questions or statements.

\vspace{0.3em}
\textbf{Response Guidelines.}
\begin{itemize}[leftmargin=1.2em, itemsep=1pt, topsep=2pt]
\item Answer based on your implicit requirements if the question is relevant.
\item If the question is not relevant, politely indicate that you do not care about it.
\item Keep responses brief and concise (1--3 sentences).
\item Maintain a natural and conversational tone.
\item Adjust your response according to the full dialogue context.
\end{itemize}

\vspace{0.3em}
\textbf{Output Format.}
Return a JSON object:
\begin{verbatim}
{"response": "<answer>"}
\end{verbatim}

\textbf{Input Variables.}

\textbf{Conversation History:} \texttt{\{conversation\_history\}}  \\
\textbf{Interviewer's Latest Utterance:} \texttt{\{latest\_utterance\}}  \\
\textbf{Context Metadata:} \\
\hspace*{1em}Action type: \texttt{\{action\_type\}}  \\
\hspace*{1em}Relevance flag: \texttt{\{is\_relevant\}}  \\
\hspace*{1em}Relevant implicit requirement: \texttt{\{relevant\_requirement\}} (null if not relevant)

\end{tcolorbox}
\caption{Prompt template used to instantiate the oracle user in \ours{}.}
\label{fig:oracle_prompt}
\end{figure}

\subsubsection{Validation.}

To validate whether the constructed LLM-based oracle user faithfully reflects real user behavior in conversational requirements elicitation, we conduct a comparative study between oracle user responses and real user responses. 

\textbf{Validation Design.} \textbf{(1) Data Selection.} We adopt two real-world interview scenarios previously used by Ferrari et al.~\cite{ferrari2020sapeer}: a hair and nail salon (Salon) seeking a digital solution for appointment booking and employee scheduling, and a ski resort chain (Ski) requiring an online booking and business management platform for three geographically distributed resorts. They contain a total of 33 interview sessions with real users, comprising 555 interviewer turns. \textbf{(2) Validation Setup.} For each interview session, we replay the interviewer’s utterances turn by turn and replace the real user with our oracle user. For each sampled turn, the oracle user generates a response conditioned on the full dialogue history and the annotated relevance of the interviewer’s question.
We then compare the oracle-generated responses with the real user responses under identical interviewer utterances. This comparison focuses on two key aspects of disclosure behavior. First, when the real user does not reveal implicit requirements, the oracle user should also refrain from revealing additional requirements, thereby avoiding unnecessary or excessive disclosure. Second, when the real user reveals requirements, the oracle user is expected to provide corresponding information in a manner consistent with the real user. \textbf{(3) Validation Metrics.}  Based on the above comparison, we evaluate the agreement between oracle user responses and real user responses across the 555 dialogue turns. Specifically, we compute Cohen’s $\kappa$ to measure overall behavioral consistency, together with the false positive (FP) rate and false negative (FN) rate. Here, FP rate reflects cases where the oracle user discloses requirement-related information that real users would not reveal under the same context, while FN rate reflects cases where the oracle user fails to disclose information that real users would provide.

\begin{table}[h]
\centering
\setlength{\tabcolsep}{6pt}
\caption{Oracle user validation on disclosure behavior.
We report mean $\pm$ standard deviation across multiple interview sessions for each system.
\#Interviews and \#Turns denote the number of interview sessions and interviewer turns, respectively.
Cohen’s $\kappa$ measures agreement with human annotations on whether an implicit requirement should be disclosed.
FP rate denotes premature disclosure (disclose when it should not), and FN rate denotes missed disclosure (fail to disclose when it should).}
\label{tab:oracle_user_validation}
\begin{tabular}{lccccc}
\toprule
\textbf{System}
& \textbf{\#Interviews}
& \textbf{\#Turns}
& $\boldsymbol{\kappa}$
& \textbf{FP rate} $\downarrow$
& \textbf{FN rate} $\downarrow$ \\
\midrule
Salon
& 15
& 233
& $0.78 \pm 0.09$
& $0.06 \pm 0.04$
& $0.11 \pm 0.07$ \\
Ski
& 18
& 322
& $0.69 \pm 0.12$
& $0.09 \pm 0.06$
& $0.15 \pm 0.08$ \\
\midrule
\textbf{Total}
& \textbf{33}
& \textbf{555}
& $0.73$
& $0.08$
& $0.13$ \\
\bottomrule
\end{tabular}
\end{table}

\textbf{Validation Results.}
Table~\ref{tab:oracle_user_validation} reports the validation results of the oracle user on disclosure behavior. Overall, the oracle user shows strong behavioral consistency with real users, achieving Cohen’s $\kappa$ scores of 0.78 and 0.69 in the Salon and Ski scenarios, respectively, and an overall agreement of 0.73. Besides, the oracle user maintains low FP rates, indicating that it rarely discloses requirement information when real users would not do so.
At the same time, the FN rates remain moderate, suggesting that the oracle user generally reveals relevant information when appropriate.
These results support the use of the oracle user as a reliable and reproducible proxy for real users in conversational requirements elicitation evaluation.

\begin{figure}[t]
\centering
\begin{tcolorbox}[
    sharp corners,
    width=0.95\textwidth,
    colback=gray!5,
    colframe=black,
    boxrule=0.6pt,
    fontupper=\footnotesize,
    title=\textbf{Validation Judge Prompt Design},
    before skip=4pt,
    after skip=0pt
]

\textbf{System Role.}
You are an expert evaluator assessing the interviewer’s action in a conversational requirements elicitation process. 
Classify the interviewer’s latest action and determine whether it targets remaining implicit requirements.

\vspace{0.3em}
\textbf{Action Definitions.}
\begin{itemize}[leftmargin=1.2em, itemsep=1pt, topsep=2pt]
\item \textbf{clarify}: Ask for clarification of previously mentioned information.
\item \textbf{probe}: Explore new or more detailed requirement information.
\item \textbf{finish}: Conclude the conversation after sufficient information is collected.
\end{itemize}

\vspace{0.3em}
\textbf{Evaluation Guidelines.}
\begin{itemize}[leftmargin=1.2em, itemsep=1pt, topsep=2pt]
\item Analyze the latest utterance based on full context.
\item Determine the action type (\texttt{clarify}, \texttt{probe}, \texttt{finish}).
\item If \texttt{probe}, assess whether it targets remaining implicit requirements.
\item Provide a brief justification.
\end{itemize}

\vspace{0.3em}
\textbf{Output Format.}
Return a JSON object:
\begin{verbatim}
{"action_type": "<type>",
 "is_relevant_to_implicit_requirements": <true/false>,
 "relevant_implicit_requirement_id": "<id or null>",
 "reasoning": "<explanation>"}
\end{verbatim}

\textbf{Input Variables.}

\textbf{Initial Requirements:} \texttt{\{initial\_requirements\}}  \\
\textbf{Conversation History:} \texttt{\{conversation\_history\}}  \\
\textbf{Interviewer's Latest Utterance:} \texttt{\{latest\_utterance\}}  \\
\textbf{Remaining Implicit Requirements:} \texttt{\{remaining\_requirements\}}

\end{tcolorbox}
\caption{Prompt template used to instantiate the validation judge in \ours{}.}
\label{fig:judge_prompt}
\end{figure}

\subsection{Task Evaluator}
\FloatBarrier
\subsubsection{Construction.}

To enable automated and reproducible evaluation of interviewer behavior in conversational requirements elicitation, \ours{} introduces an LLM-based task evaluator. The task evaluator serves as an expert judge that analyzes interviewer utterances at each dialogue turn and produces structured judgments aligned with human expert annotations. Specifically, the task evaluator is designed to output two core judgments for each interviewer's turn. First, it classifies the interviewer’s action into one of three predefined interview strategies: \textit{clarify}, \textit{probe}, or \textit{finish}.
Second, it determines whether the interviewer’s utterance is relevant to any remaining implied requirements, indicating whether the turn is addressing unmet requirement information.

The construction of the task evaluator follows a prompt-based design, as illustrated in Figure~\ref{fig:judge_prompt}.
The system prompt explicitly defines the semantics of each action type and provides clear judgment criteria, while the user prompt injects structured contextual information, including the initial requirements, full conversation history, the interviewer’s latest utterance, and the set of remaining implied requirements.
This design ensures that the evaluator’s judgments are grounded in the dialogue context and aligned with the annotated requirement state, supporting consistent and reproducible evaluation across different interviews.

\begin{table}[t]
\centering
\setlength{\tabcolsep}{6pt}
\caption{Validation results of the task evaluator on real-world interview dialogues.
We report mean $\pm$ standard deviation across multiple interview sessions for each system.
\#Interviews and \#Turns denote the number of interview sessions and interviewer turns, respectively.
Cohen’s $\kappa$ measures agreement with human annotations.
FP rate denotes false positives (premature hits), and FN rate denotes false negatives (missed hits).}
\label{tab:oracle_validation}
\begin{tabular}{lcccc}
\toprule
\textbf{System} 
& $\boldsymbol{\kappa}_{\text{Action}}$
& $\boldsymbol{\kappa}_{\text{Relevance}}$
& \textbf{FP rate} $\downarrow$
& \textbf{FN rate} $\downarrow$ \\
\midrule
Salon 
& $0.83 \pm 0.21$ & $0.72 \pm 0.17$& $0.00 \pm 0.00$ & $0.09 \pm 0.07$ \\
Ski 
& $0.72 \pm 0.22$ & $0.72 \pm 0.12$& $0.00 \pm 0.00$ & $0.09 \pm 0.10$ \\
\midrule
\textbf{Total}
& \textbf{0.78}
& \textbf{0.73}
& \textbf{0.00}
& \textbf{0.09} \\
\bottomrule
\end{tabular}
\end{table}

\subsubsection{Validation.}

We validate the task evaluator by assessing its agreement with human expert annotations on real-world interview dialogues.
The evaluation is conducted on the same two systems used in the oracle user validation, namely Salon and Ski. For validation, two aspects are manually annotated by human experts for each interviewer turn:
(1) the interview strategy adopted by the interviewer, and
(2) whether the interviewer’s utterance is relevant to any implied requirements. We then apply the task evaluator to each turn and compare its outputs with the corresponding human annotations. To quantify the agreement between the task evaluator and human experts, we compute Cohen’s $\kappa$ for both action classification and relevance judgment. In addition, we report the false positive (FP) rate and false negative (FN) rate for relevance judgments, where FP indicates cases of premature relevance detection, and FN indicates missed relevance.

\textbf{Validation Results.}
Table~\ref{tab:oracle_validation} reports the validation results of the task evaluator.
Across both systems, the task evaluator achieves substantial agreement with human annotations. In particular, the evaluator demonstrates strong consistency in action classification, with $\kappa_{\text{Action}}$ scores of 0.83 for Salon and 0.72 for Ski.
For relevance judgment, the evaluator achieves comparable agreement across the two systems, with $\kappa_{\text{Relevance}}$ scores of 0.72 in both cases. Overall, these results support the reliability of the task evaluator as an automated substitute for human judgment in evaluating interviewer behavior within \ours{}.
% Moreover, the FP rates remain consistently low, indicating that the task evaluator rarely produces premature relevance judgments.
% The FN rates are moderate, suggesting that the evaluator is generally effective at identifying requirement-relevant turns, even in more complex scenarios.
Overall, these results support the reliability of the task evaluator as an automated substitute for human judgment in evaluating interviewer behavior within \ours{}.

\subsection{Implementation Details}
\ours{} is implemented in Python with a unified configuration interface to ensure reproducibility. 
Both the oracle user and the task evaluator are instantiated via API access to \texttt{GPT-5.2}. 
The oracle user is configured with a temperature of 0.7 to simulate natural stakeholder variability, while the task evaluator adopts a deterministic setting with temperature 0.0 to ensure stable and reproducible judgments. 
For both modules, the maximum number of generated tokens is set to 1024 with a timeout of 30 seconds. 
Each interview session is limited to 20 dialogue turns, and all conversation histories and evaluation results are automatically logged for subsequent analysis and verification. 
All configuration parameters are publicly released together with the source code to facilitate replication and future benchmarking.

\section{Empirical Study}
\label{sec:empirical}
To assess the interview competence of the current mainstream LLMs in conversational requirements elicitation, we conduct a systematic empirical study using \ours{}. In this section, we describe the details of our empirical study, including research questions, evaluated LLMs, metrics, and experimental setup.

\subsection{Research Questions}

Here, we introduce four research questions (RQs) investigated in our empirical study and describe the corresponding evaluation methodology for each RQ.

\textbf{RQ1: What is the overall interview competence of mainstream LLMs in conversational requirements elicitation?} This RQ evaluates the overall ability of LLMs to conduct multi-turn requirements interviews, \ie uncover implicit requirements from underspecified initial requirements. To answer this RQ, we employ six mainstream LLMs to interact with \ours{}, performing all 101 scenarios in our evaluation dataset. For each LLM, we evaluate it under two inference settings: \textit{non-CoT} and \textit{CoT}. The \textit{non-CoT} setting requires the LLMs to directly output a question or terminate the interview without producing intermediate reasoning, while the \textit{CoT} setting allows the LLMs to explicitly generate step-by-step reasoning before producing its final question. Based on the resulting dialogue process, we compute two core metrics defined in Section~\ref{}, including Implicit Requirement Elicitation Rate (IRE) and the Turn-discounted Key Question Rate (TKQR). We report the average metric scores across all scenarios as the overall interview competence of each LLM.

\textbf{RQ2: How well do LLMs apply clarification and probing strategies during conversational requirements elicitation?} 
This RQ examines the elicitation process of LLMs, focusing on whether they can appropriately apply \textit{clarification} questions to resolve explicit ambiguities and \textit{probing} questions to proactively uncover missing requirement dimensions. 
To answer this RQ, we leverage the task evaluator to classify each interviewer's turn into one of the predefined strategies. 
We then analyze the strategy distribution of each LLM across all interaction turns to assess its preference and balance between clarification and probing behaviors. In addition, we compute the \emph{Effective Strategy Ratio} (ESR) for both \textit{clarify} and \textit{probe}, defined as the proportion of turns that successfully elicit at least one implicit requirement under each strategy. 

% This allows us to quantitatively assess whether LLMs can effectively apply clarification and probing strategies, and whether their probing questions are truly helpful for uncovering implicit requirement information.

\textbf{RQ3: How does the interview competence of LLMs vary across different types of implicit requirements?} This RQ investigates whether LLMs exhibit uneven capability in eliciting different categories of implicit requirements.
In our dataset, implicit requirements are categorized into three dimensions: \textit{Interaction}, \textit{Content}, and \textit{Style}. To answer this RQ, we compute category-wise implicit requirement elicitation rates, denoted as $\mathrm{IRE}_{Int}$, $\mathrm{IRE}_{Con}$, and $\mathrm{IRE}_{Sty}$, by measuring the proportion of elicited implicit requirements within each category. We further compare the elicitation performance gap across categories to identify which requirement types are easier or harder for LLMs to uncover. This analysis helps reveal the structural weaknesses of current LLMs in requirements elicitation. 

\textbf{RQ4: How does the interview competence of LLMs vary across different application types?} This RQ investigates whether the interview competence of LLMs generalizes consistently across different web application domains. 
Our evaluation dataset spans 20 application types, covering diverse scenarios such as e-commerce websites and dashboards. 
To answer this RQ, we group all evaluation tasks according to their application types and compute the average performance metrics for each group. Specifically, we report the application-wise IRE and TKQR, and compare the relative performance ranking of different LLMs across domains. 
This analysis allows us to provide insights into the robustness and domain sensitivity of LLM-based interview competence.

\subsection{Evaluated LLMs}

We evaluate a diverse set of representative LLMs covering both closed-source and open-source models, reflecting the current landscape of state-of-the-art general-purpose language models.

\begin{itemize}
    \item \textbf{GPT-5.2} (OpenAI, closed-source): 
    a state-of-the-art general-purpose LLM with strong reasoning and instruction-following capabilities, commonly used as a baseline for complex multi-turn interactive tasks.

    \item \textbf{Claude Opus 4.5} (Anthropic, closed-source): 
    a large-scale conversational model optimized for long-context understanding and structured dialogue.

    \item \textbf{Gemini 3 Flash} (Google, closed-source): 
    a lightweight yet capable model designed for fast response and efficient interaction.

    \item \textbf{DeepSeek V3.2} (DeepSeek, open-source): 
    a recent open-source LLM with competitive reasoning performance, enabling transparent and reproducible evaluation.

    \item \textbf{Kimi K2.5} (Moonshot AI, open-source): 
    an open-source conversational model with strong natural language understanding for dialogue-centric tasks.

    \item \textbf{GLM-4.7} (Zhipu AI, closed-source): a powerful general-purpose LLM with strong bilingual instruction-following ability and robust multi-turn dialogue performance. 

    \item \textbf{Qwen3 235B A22B 2507} (Alibaba, open-source): 
    a large-scale mixture-of-experts LLM representing the upper end of open-source model capacity.
\end{itemize}

\subsection{Evaluation Metrics} \label{subsec:metrics}
We evaluate interviewer performance using three complementary measures. Among them, Implicit Requirements Elicitation Ratio (IRE) and Effective Strategy Ratio (ESR) are proposed in this work, while Turn-discounted Key Question Rate (TKQR) is adopted from prior work on the code generation task~\cite{replace}. IRE measures the overall coverage of implicit requirements achieved by the interviewer. ESR measures how effectively the interviewer applies requirements elicitation strategies during the elicitation process. TKQR measures the efficiency with which key elicitation questions are asked, emphasizing their ordering and timeliness. We detail each metric below.

\subsubsection{Implicit Requirements Elicitation Ratio (IRE)}

Let $\mathcal{R}$ denote the set of ground-truth implicit requirements for a scenario, and let $\hat{\mathcal{R}}_{\le T}$ denote the set of implicit requirements elicited by the interviewer when the interaction ends at turn $T$.
We define:
\begin{equation}
\mathrm{IRE} = \frac{|\hat{\mathcal{R}}_{\le T}|}{|\mathcal{R}|}.
\end{equation}
IRE measures the overall coverage of missing information required to transform the initial requirement into the final specification. We can further compute dimension-specific elicitation ratios for \textit{interaction}, \textit{content}, and \textit{style} requirements.

\subsubsection{Effective Strategy Ratio (ESR)}
We introduce \emph{Effective Strategy Ratio (ESR)} to measure how effectively an interviewer applies requirements elicitation strategies during multi-turn dialogue process. Using the task evaluator in \ours{}, each interviewer turn is classified as \textit{clarify}, \textit{probe}, or \textit{finish}, and judged for relevance to remaining implicit requirements.
A clarification or probing turn is considered effective if it elicits at least one implicit requirement. Let $\mathcal{T}_{s}$ denote the set of turns adopting strategy $s \in \{\textsc{Clarify}, \textsc{Probe}\}$, and let $\mathcal{T}_{s}^{+} \subseteq \mathcal{T}_{s}$ denote the subset of effective turns.
We define:
\begin{equation}
\label{eq:esr}
\mathrm{ESR}_{s} = \frac{|\mathcal{T}_{s}^{+}|}{|\mathcal{T}_{s}|}.
\end{equation}

$\mathrm{ESR}_{s} \in [0,1]$ indicates the proportion of effective turns under strategy $s$. Higher values indicate that the interviewer can apply the strategy $s$ to elicit missing requirements better. 

\subsubsection{Turn-discounted Key Question Rate (TKQR)}
TKQR is adapted from normalized Discounted Cumulative Gain (nDCG) in information retrieval, and has been adopted in prior work on the code generation task. It rewards asking key clarification questions early and penalizes delayed or redundant questioning. Let $n$ denote the number of dialogue turns before the interviewer stops asking questions, and let $K$ denote the number of annotated implicit requirements for the scenario. We first compute a hit indicator sequence $H=(h_1,\dots,h_n)$, where $h_i=1$ if the interviewer asks a previously predefined implicit requirement at turn $i$, and $h_i=0$ otherwise. Then TKQR defines the discounted cumulative gain:
\begin{equation}
\mathrm{DCG}_n = \sum_{i=1}^{n} \frac{h_i}{\log_2(i+1)}.
\end{equation}
To normalize across scenarios with different numbers of implicit requirements, TKQR computes the ideal discounted gain:
\begin{equation}
\mathrm{IDCG}_n = \sum_{i=1}^{\min(n,K)} \frac{1}{\log_2(i+1)}.
\end{equation}
Finally, TKQR is defined as:
\begin{equation}
\mathrm{TKQR} = \frac{\mathrm{DCG}_n}{\mathrm{IDCG}_n}.
\end{equation}
TKQR ranges in $[0,1]$, where higher values indicate that the interviewer prioritizes key elicitation questions earlier in the interaction.
% Turn-discounted Key Question Rate (TKQR) is adapted from the normalized Discounted Cumulative Gain (nDCG) metric originally proposed in information retrieval and has been adopted in prior work on clarification question evaluation.
% It rewards asking key clarification questions early in the interaction and penalizes delayed or redundant questioning.
% In this study, TKQR is used to assess how efficiently an interviewer prioritizes critical missing information during requirements elicitation.

\subsection{Experimental Setup}
We describe 
% the experimental setup of our empirical study, including 
the interviewer prompt design and LLM inference settings.

\textbf{Interviewer Prompt.} For all evaluated LLMs, we assign them the role of an interviewer in conversational requirements elicitation. The system prompt explicitly constrains the model behavior: at each turn, the interviewer must either ask exactly one question to elicit missing requirements or terminate the interview by outputting a special finish signal. 
This design prevents the LLM from asking multiple questions within a single turn, thereby ensuring consistent turn granularity across different LLMs. 
To ensure a fair comparison, all models are evaluated using the same interviewer prompt. The full prompt is provided in our replication package.

\textbf{LLM Inference Setting.} We evaluate each LLM under two inference settings: \textit{non-CoT} and \textit{CoT}.
In the \textit{non-CoT} setting, the model is instructed to directly output a question (or finish decision) without producing intermediate reasoning.
In the \textit{CoT} setting, the model is allowed to generate step-by-step reasoning before producing its final question or finish decision.
For both settings, we set the decoding temperature to 0 to reduce randomness and improve reproducibility.
We set the maximum output length to 2048 tokens for \textit{non-CoT}. 
For \textit{CoT}, to avoid truncating long reasoning traces, we set the maximum output length to 8196 tokens.

\section{Experimental Results and Analysis}

\subsection{RQ1: What is the interview competence of current mainstream LLMs in conversational requirements elicitation?}
Table~\ref{tab:rq1} demonstrates the evaluation results of each model on two inference settings in terms of the number of turns, IRE, and TKQR. \textbf{(1) Current mainstream LLMs still exhibit limited capability in uncovering implicit requirements, as reflected by the generally low IRE across all models.} Even the best-performing model only achieves an IRE of 0.32 under non-CoT (\ie DeepSeek V3.2). This indicates that most models fail to systematically identify a large portion of missing requirement information from underspecified initial requirements. (2) \textbf{CoT prompting does not consistently improve elicitation coverage (IRE), but it generally enhances questioning efficiency (TKQR).} Most models show decreased IRE under CoT (\eg  DeepSeek), indicating that allowing explicit reasoning does not guarantee better elicitation. In contrast, TKQR increases for all models under CoT, demonstrating that CoT tends to encourage models to ask more relevant key questions earlier.  \textbf{(3) The number of turns reveals substantial differences in termination behaviors across LLMs.} Under the non-CoT setting, GPT-5.2 reaches an average of 19.98 turns, almost always exhausting the maximum turn budget, suggesting that it lacks an effective stopping criterion. In contrast, most other models terminate much earlier, indicating more aggressive finish decisions. Interestingly, CoT prompting substantially reduces the average turns, implying that CoT can help models make earlier termination decisions.

\begin{table}[]
    \centering
    \caption{Overall interview competence of different LLMs.
    We report mean $\pm$ standard deviation over all scenarios.}
    \begin{tabular}{lcccccc}
\toprule
\multirow{2}{*}{\textbf{LLMs}} 
& \multicolumn{2}{c}{\textbf{\#Turns}} 
& \multicolumn{2}{c}{\textbf{IRE $\uparrow$}} 
& \multicolumn{2}{c}{\textbf{TKQR $\uparrow$}} \\ 
\cmidrule(lr){2-3} \cmidrule(lr){4-5} \cmidrule(lr){6-7}
& \multicolumn{1}{c}{Non-CoT} 
& \multicolumn{1}{c}{CoT} 
& \multicolumn{1}{c}{Non-CoT} 
& \multicolumn{1}{c}{CoT} 
& \multicolumn{1}{c}{Non-CoT} 
& \multicolumn{1}{c}{CoT} \\ 
\midrule
GPT-5.2                        &  19.98 $\pm$ 0.20&  8.12 $\pm$ 1.74&  0.13 $\pm$ 0.03&  0.11 $\pm$ 0.02&  0.09 $\pm$ 0.03&  0.21 $\pm$ 0.05\\
Claude Opus 4.5                &  6.36 $\pm$ 2.26&  5.94 $\pm$ 1.88&  0.08 $\pm$ 0.02&  0.07 $\pm$ 0.02&  0.19 $\pm$ 0.04&  0.28 $\pm$ 0.06\\
Gemini 3 Flash                 &  7.76 $\pm$ 2.74&  6.85 $\pm$ 2.31&  0.11 $\pm$ 0.02&  0.10 $\pm$ 0.02&  0.11 $\pm$ 0.03&  0.18 $\pm$ 0.04\\
DeepSeek V3.2                  &  11.38 $\pm$ 3.61&  5.73 $\pm$ 1.91&  0.32 $\pm$ 0.04&  0.19 $\pm$ 0.03&  0.35 $\pm$ 0.05&  0.41 $\pm$ 0.07\\
Kimi K2.5                      &  5.76 $\pm$ 2.36&  5.46 $\pm$ 1.21&  0.19 $\pm$ 0.04&  0.20 $\pm$ 0.03&  0.33 $\pm$ 0.04&  0.55 $\pm$ 0.08\\
GLM-4.7                        &  3.86 $\pm$ 2.15&  4.53 $\pm$ 1.60&  0.15 $\pm$ 0.03&  0.20 $\pm$ 0.04&  0.24 $\pm$ 0.07&  0.33 $\pm$ 0.07\\
Qwen3 235B&  4.20 $\pm$ 0.99&  4.52 $\pm$ 1.01&  0.13 $\pm$ 0.03&  0.14 $\pm$ 0.01&  0.22 $\pm$ 0.08&  0.25 $\pm$ 0.03\\ 
\bottomrule
\end{tabular}
    \label{tab:rq1}
\end{table}

\begin{boxK}
\small \faIcon{pencil-alt} \textbf{Answer to RQ1:} Current mainstream LLMs still show limited interview competence in conversational requirements elicitation, with overall IRE remaining low across models. CoT prompting consistently improves TKQR and reduces dialogue turns, but does not reliably increase IRE, suggesting that it mainly enhances questioning efficiency rather than elicitation coverage.
\end{boxK}

\subsection{RQ2: How well do LLMs apply clarification and probing strategies during conversational requirements elicitation?}
Table~\ref{tab:rq2} reports the clarification and probing behaviors of different LLMs under non-CoT and CoT settings, including the number of clarification/probing turns and the effective strategy ratios. \textbf{(1) All LLMs strongly favor probing over clarification.} Across all models and inference settings, the number of probing turns is substantially larger than clarification turns. This indicates that current LLM-based interviewers generally prefer exploratory elicitation behaviors, while clarification questions are consistently underused. \textbf{(2) Clarification is consistently less effective than probing, indicating that LLMs struggle to ask precise clarification questions.} Under the non-CoT setting, ESR$_{clarify}$ remains low for most models. In contrast, ESR$_{prob}$ is substantially higher across all models, suggesting that probing questions are more likely to uncover implicit requirements. \textbf{(3) CoT prompting significantly improves the effectiveness of both strategies, especially probing}. Under the CoT setting, all models show consistent improvements in both ESR$_{clarify}$ and ESR$_{prob}$. For instance, GPT-5.2 improves ESR$_{prob}$ from 0.22 to 0.43. These results suggest that CoT prompting can substantially enhance per-turn strategy quality by encouraging more effective elicitation questions. However, such improvements do not necessarily translate into higher overall elicitation coverage (IRE), as CoT may lead models to terminate earlier or repeatedly elicit relatively easy requirement dimensions.

\begin{table}[]
    \centering
    \caption{Clarification and probing behavior across different inference settings.}
    \begin{tabular}{lcccccc}
\toprule
\multirow{2}{*}{\textbf{LLMs}} 
& \multicolumn{2}{c}{\textbf{\#Clarify / \#Prob}} 
& \multicolumn{2}{c}{\textbf{ESR$_{clarify}$ $\uparrow$}} 
& \multicolumn{2}{c}{\textbf{ESR$_{prob}$ $\uparrow$}} \\ 
\cmidrule(lr){2-3} \cmidrule(lr){4-5} \cmidrule(lr){6-7}
& \multicolumn{1}{c}{Non-CoT} 
& \multicolumn{1}{c}{CoT} 
& \multicolumn{1}{c}{Non-CoT} 
& \multicolumn{1}{c}{CoT} 
& \multicolumn{1}{c}{Non-CoT} 
& \multicolumn{1}{c}{CoT} \\ 
\midrule
GPT-5.2                        &  23 / 414&  31 / 203&  0.04&  0.11&  0.22&  0.43\\
Claude Opus 4.5                &  7 / 109&  15 / 82&  0.29&  0.42&  0.51&  0.61\\
Gemini 3 Flash                 &  16 / 144&  22 / 108&  0.25&  0.36&  0.47&  0.57\\
DeepSeek V3.2                  &  24 / 275 &  28 / 164 &  0.05 &  0.13 &  0.47 &  0.62 \\
Kimi K2.5                      &  19 / 238 &  25 / 176 &  0.06 &  0.14 &  0.50 &  0.68 \\
GLM-4.7                        &  15 / 142 &  22 / 128 &  0.07 &  0.16 &  0.39 &  0.52 \\
Qwen3 235B A22B 2507&  12 / 118 &  16 / 101 &  0.04 &  0.09 &  0.33 &  0.41 \\ 
\bottomrule
\end{tabular}
    \label{tab:rq2}
\end{table}
\begin{table}[t]
    \centering
    \caption{Implicit requirements elicitation ratio (IRE) across different requirement types.
    We report the proportion of elicited implicit requirements in Interaction (Int), Content (Con), and Style (Sty).}
    \label{tab:rq3}
    \begin{tabular}{lcccccc}
        \toprule
        \multirow{2}{*}{\textbf{LLMs}}  & \multicolumn{3}{c}{\textbf{Non-CoT}} & \multicolumn{3}{c}{\textbf{CoT}} \\
        \cmidrule(lr){2-4} \cmidrule(lr){5-7}
        & $\boldsymbol{\mathrm{IRE}_{Int}}$ 
        & $\boldsymbol{\mathrm{IRE}_{Con}}$ 
        & $\boldsymbol{\mathrm{IRE}_{Sty}}$ 
        & $\boldsymbol{\mathrm{IRE}_{Int}}$ 
        & $\boldsymbol{\mathrm{IRE}_{Con}}$ 
        & $\boldsymbol{\mathrm{IRE}_{Sty}}$ \\
        \midrule
        GPT-5.2              & 0.19& 0.13& <0.01& 0.12& 0.09& <0.01\\
        Claude Opus 4.5      & 0.08& 0.12& <0.01& 0.06& 0.09& <0.01\\
        Gemini 3 Flash       & 0.12& 0.16& <0.01& 0.09& 0.12& <0.01\\
        DeepSeek V3.2        & 0.48 & 0.46 & 0.01 & 0.29 & 0.27 & <0.01\\
        Kimi K2.5            & 0.31 & 0.23 & <0.01& 0.33 & 0.25 & <0.01\\
        GLM-4.7              & 0.21 & 0.20 & 0.01 & 0.30 & 0.26 & <0.01\\
        Qwen3 235B A22B 2507 & 0.20 & 0.19 & <0.01& 0.22 & 0.20 & <0.01\\
        \bottomrule
    \end{tabular}
\end{table}

\begin{boxK}
\small \faIcon{pencil-alt} \textbf{Answer to RQ2:} LLMs overwhelmingly favor probing over clarification during requirements interviews, while clarification questions are relatively rare and less effective. CoT prompting consistently improves the effectiveness of both strategies, especially probing, but does not necessarily increase overall elicitation coverage.
\end{boxK}

\subsection{RQ3: How does the interview competence of LLMs vary across different types of implicit requirements?}

Table~\ref{tab:rq3} reports the implicit requirement elicitation ratio (IRE) across three requirement categories, including Interaction, Content, and Style. \textbf{(1) LLMs consistently perform much better on Interaction and Content requirements than on Style requirements.}
Across all models, $\mathrm{IRE}_{Int}$ and $\mathrm{IRE}_{Con}$ are substantially higher than $\mathrm{IRE}_{Sty}$, indicating that LLMs are more capable of eliciting functional and content-related requirements than subjective preferences. \textbf{(2) LLMs largely ignore style-related implicit requirements, showing a structural weakness in uncovering subjective user preferences.} For almost all models and settings, $\mathrm{IRE}_{Sty}$ remains below 0.01, suggesting that LLM-based interviewers rarely elicit aesthetic requirements (\eg visual style).

\begin{boxK}
\small \faIcon{pencil-alt} \textbf{Answer to RQ3:} LLMs show clear disparities across implicit requirement types: Interaction and Content requirements are moderately elicitable, while Style requirements are largely ignored, with $\mathrm{IRE}_{Sty}$ remaining near zero across almost all models and settings.
\end{boxK}

\subsection{RQ4: How does the interview competence of LLMs vary across different website application types?}

\begin{figure}
    \centering
    \includegraphics[width=0.99\linewidth]{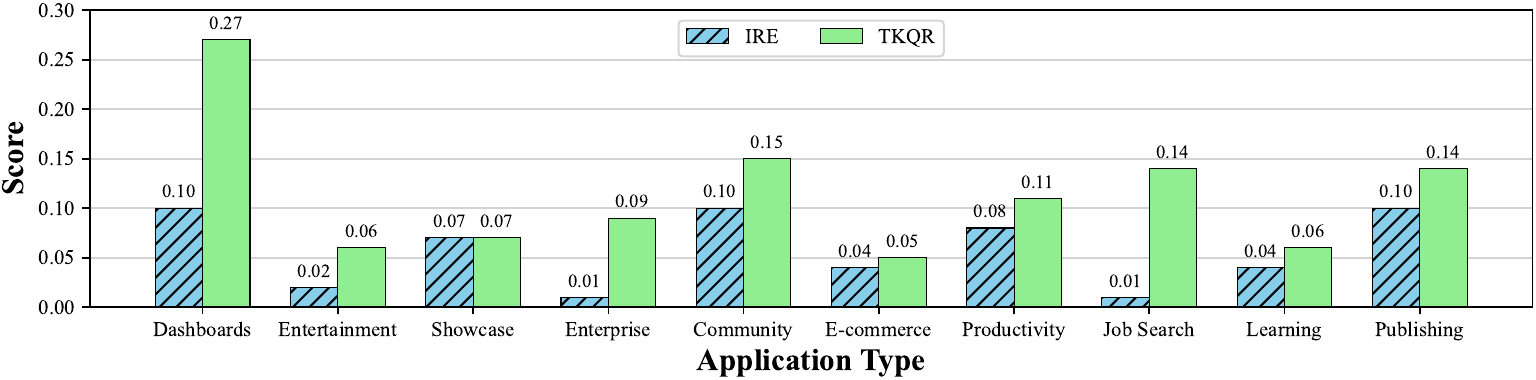}
    \caption{Average elicitation performance across different website application types. }
    \label{fig:rq4}
\end{figure}

Figure~\ref{fig:rq4} presents the average IRE and TKQR across different website application types.
\textbf{(1) Interview competence varies substantially across application types.}
IRE ranges from as low as 0.05 (E-commerce Web) and 0.06 (Learning Platforms) to as high as 0.27 (Dashboards), indicating that elicitation difficulty is highly application-type dependent. TKQR also varies across application types, suggesting that both coverage and questioning efficiency are influenced by the characteristics of each application type. \textbf{(2) Structured and task-oriented application types (\eg Dashboards and Community Platforms) achieve higher elicitation performance.} Dashboards obtain the highest IRE (\ie 0.27), followed by Community Platforms (\ie 0.15) and Publishing Platforms (\ie 0.14). These application types typically involve clearer functional structures and interaction logic, making missing requirements easier to identify and articulate.

\begin{boxK}
\small \faIcon{pencil-alt} \textbf{Answer to RQ4:} Interview competence varies notably across application types. Structured and task-oriented application types achieve higher elicitation performance. LLM-based interview remains strongly application-type dependent and lacks uniform robustness. 
\end{boxK}

\section{Discussion}
In this section, we discuss potential threats, and outline promising directions for future research.
\subsection{Threats to Validity}\label{subsec:threats}

We discuss potential threats to validity from three widely adopted perspectives, \ie construct validity, internal validity, and external validity.

\textbf{Construct Validity} concerns whether our evaluation metrics and experimental design accurately measure interview competence in conversational requirements elicitation. 
First, we acknowledge that interview competence is inherently multi-dimensional. To mitigate this threat, we adopt multiple quantitative metrics, \ie IRE (coverage), TKQR (efficiency), and ESR (strategy effectiveness), which collectively characterize both outcome-level performance and process-level behaviors. 
Second, the annotation of implicit requirements and initial requirements involves human judgment and may introduce subjectivity. To reduce this risk, we employ a structured difference-based annotation procedure (Final Spec vs. Initial Spec), follow explicit annotation guidelines, and conduct three rounds of iterative quality refinement until convergence. 
Third, both the oracle user and the task evaluator are implemented using LLM-based prompts. To ensure their reliability, we validate them against real-world interview data and expert annotations, achieving substantial agreement (Cohen’s $\kappa$ above 0.70). 
These validations provide evidence that the constructed evaluation environment reasonably approximates human behavior and expert judgment.

\textbf{Internal Validity} addresses potential threats to the way the study was conducted. The potential threats arising from experimental design choices that may influence the observed results. One potential threat stems from prompt design, as different prompts may favor certain models. To mitigate this risk, all evaluated LLMs share an identical interviewer prompt, interaction protocol, and turn limit. 
Another risk arises from inference configurations, such as reasoning settings and decoding randomness, which may affect model behavior. To control this factor, we evaluate all models under two predefined settings (non-CoT and CoT) using a unified configuration, and fix the decoding temperature to 0 to reduce stochastic variance and improve reproducibility. 
A further threat lies in inconsistencies in the evaluation pipeline. To ensure fairness, the same oracle user and task evaluator are applied uniformly across all models, and all interaction logs and configuration parameters are recorded and publicly released to facilitate replication.

\textbf{External Validity} concerns the generalizability of our findings. One potential threat arises from domain limitation. Our dataset focuses on website development scenarios. Although these scenarios are representative of many real-world software systems, they may not fully cover other domains such as embedded systems, enterprise systems, or safety-critical applications. To mitigate this risk, our dataset contains 101 scenarios spanning 10 diverse application types, aiming to cover a wide range of functional and non-functional requirements within the web domain. Nevertheless, our evaluation environment remains a controlled abstraction of real-world requirements elicitation. Future work may extend the dataset to additional domains.

\subsection{Future Directions}

Our empirical findings reveal that current LLMs still exhibit limited and structurally unbalanced interview competence. 
These suggest that future research should move beyond free-form conversational behavior and toward more structured elicitation mechanisms. We outline two promising directions for enhancement.

\textbf{(1) Ontology-Guided Structured Interviewing.} One fundamental limitation of current LLM-based interviewers is the absence of an explicit and structured representation of requirement dimensions.  Future work may integrate ontology engineering techniques to construct domain-specific requirement ontologies from large corpora of requirements artifacts. Such ontologies can explicitly model requirement categories, sub-dimensions, dependencies, and constraints, forming a structured inquiry space. During conversation, the LLM can leverage the ontology as a reasoning scaffold, systematically traversing requirement dimensions rather than relying on implicit pattern completion. 
This approach may help reduce dimension omission and encourage more comprehensive coverage of implicit information.

\textbf{(2) Reinforcement Learning for Strategy Optimization.}
Our results show that LLMs overwhelmingly favor probing over clarification and often lack effective stopping criteria. 
This suggests that conversational requirements elicitation should be formulated as a sequential decision-making problem. Future research may apply agentic reinforcement learning (RL) to train LLM-based interviewer agents, where actions correspond to strategy selection (\eg clarify, probe, finish), and rewards are defined based on coverage (IRE), efficiency (TKQR), and strategy effectiveness (ESR). 
Such training could teach models not only \emph{what} to ask, but also \emph{when} and \emph{how} to ask it.

\section{Conclusion}
This paper investigates an increasingly critical yet underexplored problem in the era of LLM-based automated software development: the interview competence of LLMs in conversational requirements elicitation. Centered on this problem, we propose an automated and reproducible evaluation environment, \ours{}, for conversational requirements elicitation. Using \ours{}, any automated elicitation approach can be evaluated in a reproducible and quantitative manner through interaction with the environment. Based on this environment, we conduct a comprehensive evaluation of seven representative LLMs. The results show that current models still exhibit limited coverage of implicit requirements, highlighting a clear gap between conversational fluency and systematic elicitation capability. Finally, we outline two promising research directions for advancing LLM-based conversational requirements elicitation.

% \section{Data Availability}
% Our source code, evaluation dataset, and experimental data are available at \url{https://github.com/jdm4pku/ReqElicitBench.}
% \input{chapters/9-env}

% 加一些过程性的指标。

%\bibliographystyle{unsrt}
\bibliographystyle{ACM-Reference-Format}
\bibliography{myrefs}

\end{document}